\documentclass{elsart}

\usepackage{amsmath,amsfonts,amssymb}
\usepackage{mathrsfs,bbm}                               
\usepackage{graphicx,subfigure,
,psfrag}					

\usepackage[sort&compress]{natbib}			
\bibpunct{[}{]}{,}{n}{}{;}				

	\newcommand{\abs}[1]{\ensuremath{\left|#1\right|}}

	\renewcommand{\Re}{\ensuremath{\mathbb{R}}}
	\newcommand{\N}{\ensuremath{\mathbb N}}

	\DeclareMathAlphabet{\mathpzc}{OT1}{pzc}{m}{it}
\begin{document}
\begin{frontmatter}
\title{Wavelet entropy of stochastic processes}
\author[ciop,fis]{L. Zunino\corauthref{cor}},
\corauth[cor]{Corresponding author.}
\ead{lucianoz@ciop.unlp.edu.ar}
\author[pucv]{D. G. P\'erez},
\ead{dario.perez@ucv.cl}
\author[ciop,fis]{M. Garavaglia}
\ead{garavagliam@ciop.unlp.edu.ar}
and
\author[uba]{O. A.  Rosso}
\ead{oarosso@fibertel.com.ar}

\address[ciop]{Centro de Investigaciones \'Opticas (CIOp), \\ CC. 124 Correo Central,1900 La Plata, Argentina.}

\address[fis]{Departamento de F\'{\i}sica, Facultad de Ciencias Exactas, \\ Universidad Nacional de La Plata (UNLP), 1900 La Plata, Argentina.}

\address[pucv]{Instituto de F\'isica, Pontificia Universidad Cat\'olica de Valpara\'iso (PUCV), 23-40025 Valpara\'iso, Chile.}

\address[uba]{Instituto de C\'alculo, Facultad de Ciencias Exactas y Naturales, \\ Universidad de Buenos Aires (UBA), Pabell\'on II, Ciudad Universitaria, 1428 Ciudad de Buenos Aires, Argentina. }


\begin{abstract} 
We compare two different definitions for the wavelet entropy associated to  stochastic processes. The first one, the Normalized Total Wavelet Entropy (NTWS) family [Phys. Rev. E \textbf{57} (1998) 932; J. Neuroscience Method \textbf{105} (2001) 65; Physica A (2005) in press] and a second introduced by Tavares and Lucena [Physica A \textbf{357} (2005)~71].
In order to understand their advantages and disadvantages, exact results obtained for fractional Gaussian noise ($-1<\alpha<\ 1$) and the fractional Brownian motion ($1 <\alpha<\ 3$) are assessed. 
We find out that NTWS family performs better as a characterization method for these stochastic processes.
\end{abstract}

\end{frontmatter}

\section{Introduction}\label{sec:intro}

The advantages of projecting an arbitrary continuous stochastic process in a discrete wavelet space are widely known. The wavelet time-frequency representation does not make any assumptions about signal stationarity and is capable of detecting dynamic changes due to its localization properties~\cite{book:daubechies92,book:mallat1999}. Unlike the harmonic base functions of the Fourier analysis, which are precisely localized in frequency but infinitely extend in time, wavelets are well localized in both time and frequency. Moreover, the computational time is significantly shorter since the algorithm involves the use of fast wavelet transform in a multiresolution framework~\cite{book:mallat1999}. Finally, contaminating noises' contributions can be easily eliminated when they are concentrated in some frequency bands. These important reasons justify the introduction, within this special space, of entropy-based algorithms in order to quantify the degree of order-disorder associated with a multi-frequency signal response. With the entropy estimated via the wavelet transform, the time evolution of frequency patterns can be followed with an optimal time-frequency resolution. In this paper we focus on two definitions for this quantifier: the Normalized Total Wavelet Entropy (NTWS) family introduced recently by us~\cite{paper:blanco98,paper:rosso2001,paper:zunino2004,paper:perez2005}, and another definition given by Tavares and Lucena~\cite{paper:tavares2005}. We compare their performance while characterizing two important stochastic processes: the fractional Brownian motion (fBm) and the fractional Gaussian noise (fGn). In particular, we will show that the NTWS family gives a better characterization for both of them.

\section{Wavelet quantifiers}\label{sec:wq}
\subsection{Wavelet energies}\label{ssec:we}
The Wavelet Analysis is one of the most useful tools when dealing with data samples. Any signal can be decomposed by using a wavelet dyadic discrete family $\{ 2^{j/2}\psi(2^j t -k) \}$, with $j,k\in\mathbb{Z}$ (the set of integers)---an  \textit{orthonormal} basis for $L^2(\mathbb{R})$ consisting of finite-energy signals---of translations and scaling functions based on a function $\psi$: the mother wavelet~\cite{book:daubechies92,book:mallat1999}. In the following, given a stochastic process $s(t)$ its associated \textit{signal} is assumed to be given by the sampled values $\mathcal{S}=\{s(n), n=1,\cdots,M\}$. Its wavelet expansion has associated wavelet coefficients given by 
\begin{equation}
C_j(k) = \langle \mathcal{S}, 2^{j/2}\psi(2^j \cdot -k) \rangle,
\end{equation}
with $j=-N,\cdots,-1$, and $N = \log_2 M$. The number of coefficients at each resolution level is $N_j=2^j M$. Note that this correlation gives information on the signal at scale $2^{-j}$ and time $2^{-j}k$. The set of wavelet coefficients at level $j$, $\{C_j(k)\}_{k}$, is also a stochastic process where $k$ represents the discrete time variable. It provides a direct estimation of local energies at different scales. Inspired by the Fourier analysis we define the energy at resolution level $j$ by
\begin{equation}
\mathcal{E}_j = \sum_k \mathbb{E}\abs{C_j(k)}^2,
\end{equation}
where $\mathbb{E}$ stands for the average using some, at first, unknown probability distribution. In the case the set $\{C_j(k)\}_k$ is proved to be a stationary process the previous equation reads
\begin{equation}
\mathcal{E}_j = N_j \mathbb{E}\abs{C_j(k)}^2.
\label{eq:ew}
\end{equation}
Observe that the energy $\mathcal{E}_j $ is only a function of the resolution level. Also, under the same assumptions, the temporal average energy at level $j$ is given by
\begin{equation}
\widetilde{\mathcal{E}}_j = \frac{1}{N_j} \sum_k \mathbb{E}\abs{C_j(k)}^2 =  \mathbb{E}\abs{C_j(k)}^2,
\end{equation}
where we have used eq.~\eqref{eq:ew} to arrive to the last step in this equation. Since we are using dyadic discrete wavelets the number of coefficients decreases over the low frequency bands (at resolution level $j$ the number is halved with respect to the previous one $j+1$); thus, the latter energy definition reinforce the contribution of these low frequency bands.

Summing over all the available wavelets levels $j$ we obtain the corresponding total energies:
$\mathcal{E}_\text{total} = \sum_{j=-N}^{-1}\mathcal{E}_j$ and $\widetilde{\mathcal{E}}_\text{total} = \sum_{j=-N}^{-1}\widetilde{\mathcal{E}}_j$. Finally, we define the \textit{relative wavelet energy}
\begin{equation}
p_j=\frac{\mathcal{E}_j}{\mathcal{E}_\text{tot}},
\label{eq:rwe}
\end{equation}
and the \textit{relative temporal average wavelet energy}
\begin{equation}
\widetilde{p}_j=\frac{\widetilde{\mathcal{E}}_j}{\widetilde{\mathcal{E}}_\text{tot}}.
\label{eq:tawe}
\end{equation}
Both supply information about the relative energy associated with the different frequency bands. So, they enable us to learn about their corresponding degree of importance. Clearly, $\sum_{j=-N}^{-1} p_j=\sum_{j=-N}^{-1} \widetilde{p}_j=1$; both define probability distributions: $\{p_j\}$ and $\{\widetilde{p}_j\}$---they can also be considered as time-scale energy densities.

\subsection{Normalized Total Wavelet Entropy family}\label{ssec:ntws}

The Shannon entropy \cite{paper:shannon48} provides a measure of the information of any distribution.																																																																																				       
Consequently, we have previously defined the family of \textit{Normalized Total Wavelet Entropy} (NTWS) as~\cite{paper:blanco98,paper:rosso2001}
\begin{equation}
\label{eq:we1}
S_\text{W}(N)=  -\sum_{j=-N}^{-1}   p_j  \cdot \log_2 p_j/ S_\text{max},
\end{equation}
and,
\begin{equation}
\label{eq:we2}
\widetilde{S}_\text{W}(N)=  -\sum_{j=-N}^{-1}  \widetilde{p}_j  \cdot \log_2 \widetilde{p}_j/ S_\text{max},
\end{equation}
with $S_\text{max}=\log_2 N.$ It has been adopted the base-2 logarithm for the entropy definition to take advantage of the dyadic nature of the wavelet expansion; thus, simplifying the entropy formulae that will be used in this work.

\subsection{Tavares-Lucena Wavelet Entropy}\label{ssec:tlws}

Alternatively, Tavares and Lucena, following the basis entropy cost concept~\cite{book:mallat1999}, have recently~\cite{paper:tavares2005} defined another probability distribution:
\begin{equation}
p_{jk}=\mathbb{E}\abs{C_j(k)}^2/\mathcal{E}^\text{(TL)}_\text{tot}\quad \text{and}\quad p_{\phi}=\mathbb{E}\abs{\langle \mathcal{S}, \phi \rangle}^2/\mathcal{E}^\text{(TL)}_\text{tot},
\end{equation}
where $\phi$ is the scaling function having the properties of a smoothing kernel (see reference~\cite{paper:tavares2005} for details), and $\mathcal{E}^\text{(TL)}_\text{tot} = \sum_{j,k}\mathbb{E}\abs{C_j(k)}^2 + \mathbb{E}\abs{\langle \mathcal{S}, \phi \rangle}^2$. Therefore, they propose the following entropy 
\begin{equation}
\label{eq:we-bras}
S^{(TL)}_\text{W}(N)=  -\left(\sum_{j=-N+1}^{j=0} \sum_{k=0}^{2^{-j}-1} p_{jk} \log_2 p_{jk} + p_{\phi} \log_2 p_{\phi}\right)\big/S_\text{max}^\text{(TL)},
\end{equation}
with $S^\text{(TL)}_\text{max}=\log_2 (2^N-1)$. As a matter of comparison we have normalized this expression and it will be referred as \textit{Tavares-Lucena Wavelet Entropy} (TLWS).

It should be noted that in eqs.~\eqref{eq:we1}, \eqref{eq:we2}, and \eqref{eq:we-bras} the maximum resolution level $N$ is an experimental parameter. It appears explicitly as a direct consequence of sampling. Tavares and Lucena underlined this fact because it is not mentioned in previous approaches.

\section{Theoretical results and comparison}\label{sec:sim}

The aim of this paper is to study the performance of the wavelet entropy definitions previously given. So we analyze two well known stochastic processes, namely, the fBm and the fGn~\cite{paper:mandelbrot1968,book:taqqu94}. The energy per resolution level $j$ and sampled time $k$ has been already evaluated for the fBm~\cite{paper:flandrin1992,paper:abry2000,paper:perez2005}. But it can be extended to fGn---see the Appendix. The final form reads
\begin{equation}
\mathbbm{E}\abs{C^\alpha_j(k)}^2 = 2\, c^2_H\, 2^{-j\alpha}\int_0^{\infty} \nu^{-\alpha} \abs{\Psi(\nu)}^2\d \nu,
\label{eq:re}
\end{equation}
where $-1<\alpha<3$---by continuity we have added $\alpha=1$ but it does not belong to any existent process. It should be noted that the latter is independent of $k$.
In the following we will use this power-law behavior with different ranges for $\alpha$, for the two stochastic processes under analysis, gathering both into a unified framework. According to its values, the coefficient $\alpha$ must be attached to one of the two mentioned processes.

In order to calculate the NTWS family, the relative wavelet energy for a finite data sample is obtained from eqs.~\eqref{eq:rwe} and~\eqref{eq:re}
\begin{equation}
p_j  = 2^{-(j + 1)(\alpha-1)} \frac{1 - 2^{\alpha-1}}{1- 2^{N(\alpha-1)}}.
\end{equation}
Similarly, the relative temporal average wavelet energy---see eqs.~\eqref{eq:tawe} and~\eqref{eq:re}---gives 
\begin{equation}
\widetilde{p}_j= 2^{-(j + 1)\alpha} \frac{1 - 2^{\alpha}}{1- 2^{N\alpha}}.
\end{equation}

Consequently, the normalized total wavelet entropies can be easily obtained from eqs.~\eqref{eq:we1} and \eqref{eq:we2},
\begin{multline}
S_\text{W}(N,\alpha)= \frac{(\alpha-1)}{\log_2 N} \left[\frac{1}{1 - 2^{- (\alpha-1)}}- \frac{N}{1- 2^{-N (\alpha-1)}}  \right] \\
- \frac{1}{\log_2 N}\log_2 \left[\frac{1-2^{(\alpha-1)}}{1-2^{N (\alpha-1)}}\right]
\label{eq:ntws-entropy}
\end{multline}
and
\begin{multline}
\widetilde{S}_\text{W}(N,\alpha)= \frac{\alpha}{\log_2 N} \left[\frac{1}{1 - 2^{- \alpha}}- \frac{N}{1- 2^{-N \alpha}}  \right] \\
- \frac{1}{\log_2 N}\log_2 \left[\frac{1-2^{\alpha}}{1-2^{N \alpha}}\right].
\label{eq:ntws-entropy}
\end{multline}

For the Tavares and Lucena's approach similar steps should be followed. From the power-law behavior mentioned before a straightforward calculation yields
\begin{equation}
p_{jk} = 2^{-j\alpha} \frac{1 - 2^{\alpha+1}}{1- 2^{N(\alpha + 1)}}.
\end{equation}
Therefore, the TLWS is obtained replacing the above into eq.~\eqref{eq:we-bras},
\begin{multline}
S^\text{(TL)}_\text{W}(N,\alpha)=
\frac{\alpha}{\log_2 (2^{N}-1)} \left[\frac{1}{1 - 2^{- (\alpha+1)}}- \frac{N}{1- 2^{-N (\alpha+1)}}  \right] \\
\\
- \frac{1}{\log_2 (2^{N}-1)}\log_2 \left[\frac{1-2^{(\alpha +1)}}{1-2^{N( \alpha +1)}}\right].
\label{eq:cost-entropy}
\end{multline}

\begin{figure} 
\begin{center}
\psfrag{N}[][t]{$N$}
\psfrag{ent}[][]{\small$S_\text{W}$}
\psfrag{alpha}[][tl]{$\alpha$}
\psfrag{0}[][]{\small$0$}
\psfrag{-1}[][]{\small$-1$}
\psfrag{1}[][]{\small$1$}
\psfrag{2}[][]{\small$2$}
\psfrag{3}[][]{\small$3$}
\psfrag{8}[][]{\small$8$}
\psfrag{10}[][]{\small$10$}
\psfrag{12}[][]{\small$12$}
\psfrag{14}[][]{\small$14$}
\psfrag{16}[][]{\small$16$}
\psfrag{0.9}[][]{\small$0.9$}
\psfrag{0.2}[][]{\small$0.2$}
\psfrag{0.3}[][]{\small$0.3$}
\psfrag{0.5}[][]{\small$0.5$}
\psfrag{0.7}[][]{\small$0.7$}
\psfrag{0.4}[][]{\small$0.4$}
\psfrag{0.6}[][]{\small$0.6$}
\psfrag{0.8}[][]{\small$0.8$}
\includegraphics[width=.8\textwidth]{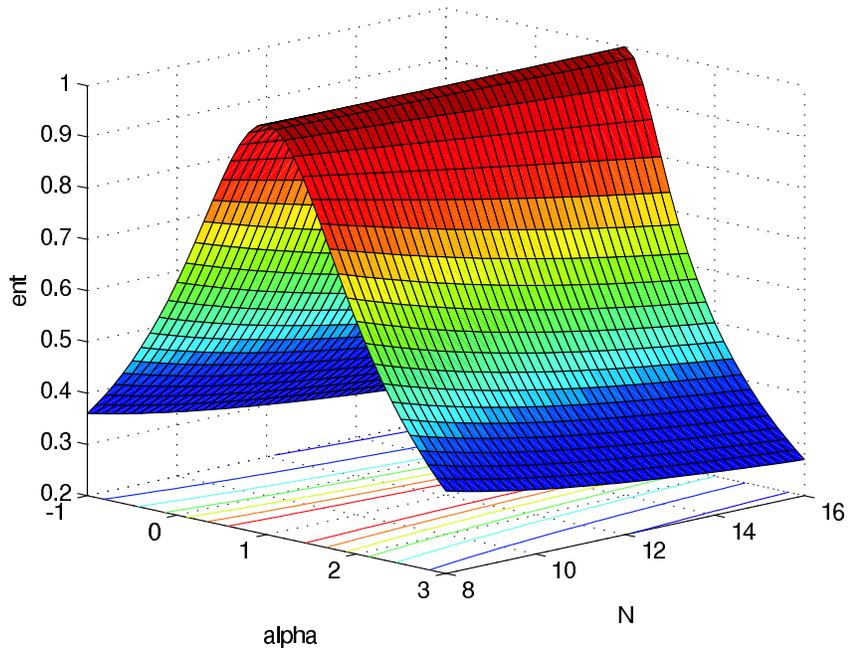}
\caption{NTWS entropy $S_\text{W}$ as a function of $\alpha$ and $N$.\label{figure:1}}
\end{center}
\end{figure}

\begin{figure} 
\begin{center}
\psfrag{N}[][t]{$N$}
\psfrag{s}[][t]{}
\psfrag{ent}[][]{\small$\widetilde{S}_\text{W}$}
\psfrag{alpha}[][tl]{$\alpha$}
\psfrag{0}[][]{\small$0$}
\psfrag{0b}[][]{\small$0$}
\psfrag{1}[][]{\small$1$}
\psfrag{-1}[][]{\small$-1$}
\psfrag{8}[][]{\small$8$}
\psfrag{2}[][]{\small$2$}
\psfrag{10}[][]{\small$10$}
\psfrag{3}[][]{\small$3$}
\psfrag{12}[][]{\small$12$}
\psfrag{14}[][]{\small$14$}
\psfrag{16}[][]{\small$16$}
\psfrag{0.2}[][]{\small$0.2$}
\psfrag{0.4}[][]{\small$0.4$}
\psfrag{0.6}[][]{\small$0.6$}
\psfrag{0.8}[][]{\small$0.8$}
\includegraphics[width=.8\textwidth]{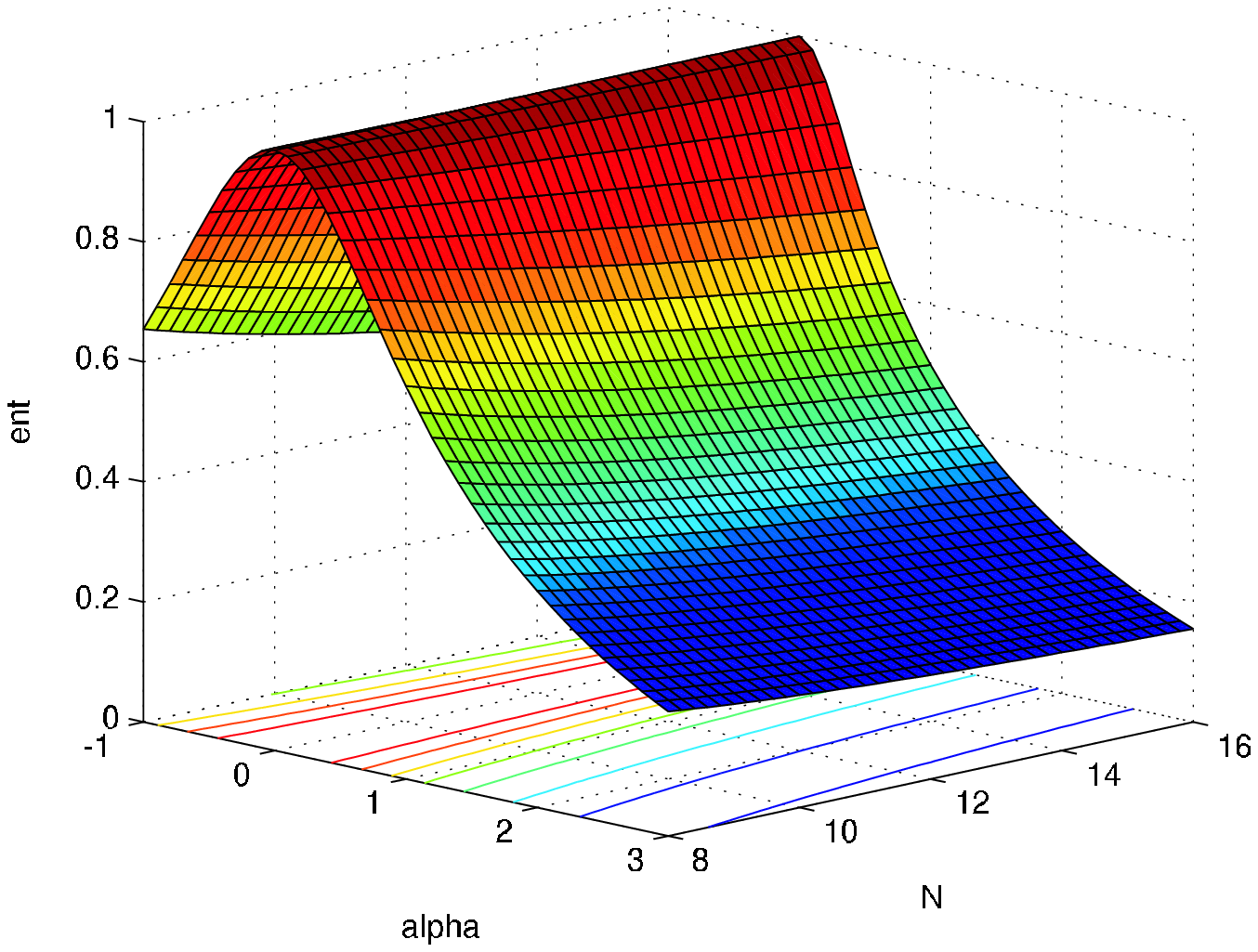}
\caption{NTWS entropy $\widetilde{S}_\text{W}$ as a function of $\alpha$ and $N$.\label{figure:2}}
\end{center}
\end{figure}

\begin{figure} 
\begin{center}
\psfrag{N}[][t]{$N$}
\psfrag{s}[][t]{}
\psfrag{ent}[l][l]{$S^\text{(TL)}_\text{W}$}
\psfrag{alpha}[][tl]{$\alpha$}
\psfrag{0}[][]{\small$0$}
\psfrag{1}[][]{\small$1$}
\psfrag{-1}[][]{\small$-1$}
\psfrag{8}[][]{\small$8$}
\psfrag{2}[][]{\small$2$}
\psfrag{10}[][]{\small$10$}
\psfrag{3}[][]{\small$3$}
\psfrag{12}[][]{\small$12$}
\psfrag{14}[][]{\small$14$}
\psfrag{16}[][]{\small$16$}
\psfrag{0.7}[][t]{\small$0.7$}
\psfrag{0.75}[][]{\small$0.75$}
\psfrag{0.85}[][]{\small$0.85$}
\psfrag{0.8}[][]{\small$0.8$}
\psfrag{0.9}[][]{\small$0.9$}
\psfrag{0.95}[][]{\small$0.95$}
\includegraphics[width=.8\textwidth]{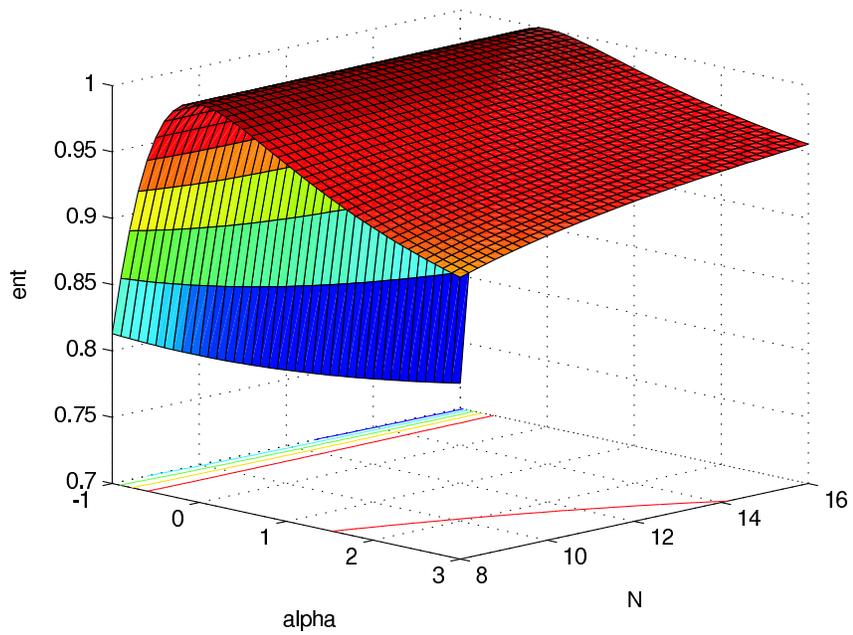}
\caption{TLWS entropy $S^\text{(TL)}_\text{W}$ as a function of $\alpha$ and $N$.\label{figure:3}}
\end{center}
\end{figure}

The NTWS family and the TLWS, as a function of $\alpha$ and $N$, are depicted in Figs.~\ref{figure:1} to \ref{figure:3}. One point to emphasize from these graphs when $\alpha >0$ is that the NTWS's range of variation increases smoothly with $N$, improving detection; on the opposite, the TLWS's range decreases when $N$ increases. All entropies equally improve with $N$ on the $-1<\alpha < 0$ branch. Moreover, for any $N$ the NTWS family covers almost all the available range between $0$ and $1$, while the TLWS roughly covers a $25\%$ of this range. 

It is of common understanding that high entropy values are associated to a signal generated by a totally disordered random process, and low values to an ordered or partially ordered process. If the process is noisy, its signal wavelet decomposition is expected to have significant contributions to the total wavelet energy coming from all frequency bands. Moreover, one could expect that all the contributions being of the same order. Consequently, its relative energies will be almost equal at all resolution levels and acquire the entropy maximum value. While a nearly ordered process will have a relative energy contribution concentrated around some level $j$, thus its entropy will take a low value. The only entropy in concordance with this intuitive vision is  $\widetilde{S}_\text{W}$, depicted in Fig.~\ref{figure:2}.

\begin{figure} 
\begin{center}
\psfrag{ant}[][t]{$S_\text{W}$}
\psfrag{nt}[][t]{$\widetilde{S}_\text{W}$}
\psfrag{tl}[][b]{$S^\text{(TL)}_\text{W}$}
\psfrag{noise}[][t]{\uppercase{\bf noise region}}
\psfrag{alpha}[][b]{$\alpha$}
\psfrag{ent}{$S$}
\psfrag{0}[][]{\small $0$}
\psfrag{1}[][]{\small$1$}
\psfrag{1.5}[][]{\small$1.5$}
\psfrag{2}[][]{\small$2$}
\psfrag{2.5}[][]{\small$2.5$}
\psfrag{3}[][]{\small$3$}
\psfrag{0.1}[][]{\small$0.1$}
\psfrag{0.2}[][]{\small$0.2$}
\psfrag{0.3}[][]{\small$0.3$}
\psfrag{0.4}[][]{\small$0.4$}
\psfrag{0.5}[][]{\small$0.5$}
\psfrag{0.6}[][]{\small$0.6$}
\psfrag{0.7}[][]{\small$0.7$}
\psfrag{0.8}[][]{\small$0.8$}
\psfrag{0.9}[][]{\small$0.9$}
\psfrag{-1}[][]{\small$-1$}
\psfrag{-0.5}[][]{\small$-0.5$}
\psfrag{0.557}[][]{\tiny$0.557$}
\psfrag{0.302}[][]{\tiny$0.302$}
\includegraphics[width=.9\textwidth]{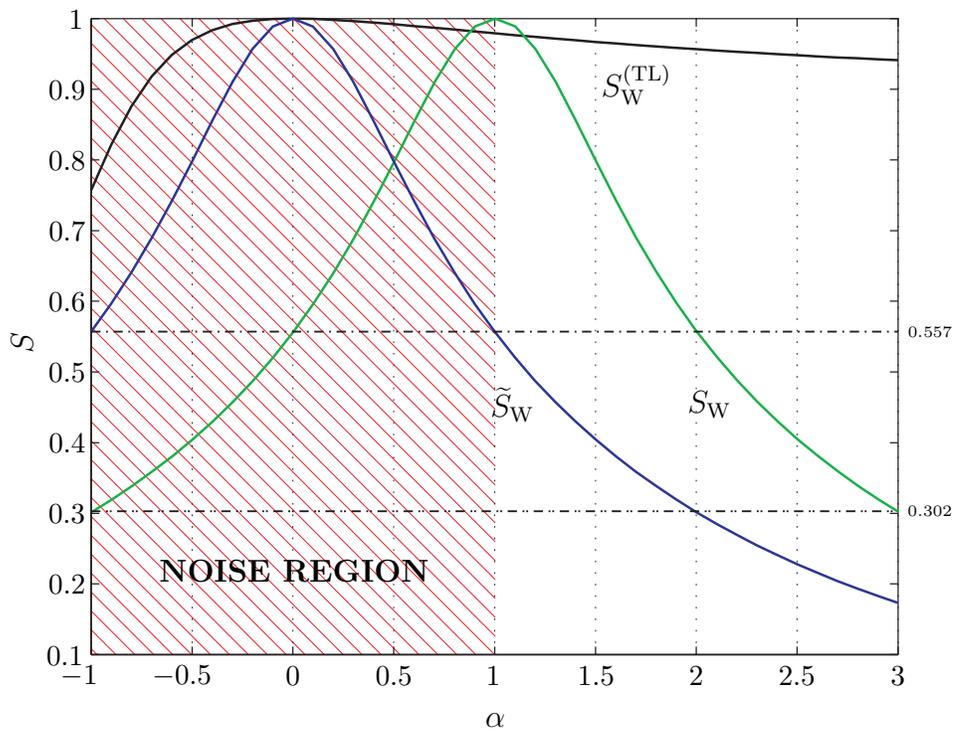}
\caption{NTWS and TLWS as functions of $\alpha$ with $N=12$.\label{figure:4}}
\end{center}
\end{figure}

In Fig.~\ref{figure:4} we compare the two entropy formulations as functions of the $\alpha$-parameter when $N=12$. It is clear that the  $\widetilde{S}_\text{W}$ and $S^\text{(TL)}_\text{W}$ entropies attain their maxima at $\alpha=0$ (white noise), and the $S_\text{W}$ entropy reaches it when $\alpha \rightarrow 1$. There are two different regions to examine: 

\begin{itemize}
\item \textit{fractional Brownian motion, $1<\alpha < 3$:} 

All the three quantifiers have their maximum at $\alpha = 1$, and monotonically decrease to find their minimum in a near regular process, $\alpha \to 3$. The range of variation of the TLWS is $\Delta S^\text{(TL)}_\text{W} = 0.038$, and the range of variation of the NTWS family is $\Delta \widetilde{S}_\text{W} = 0.384$ and $\Delta S_\text{W} = 0.698$. Clearly, due to the small range of variation, the TLWS is unfit to differentiate between the short- and long-memory fBm family members, $1<\alpha<2$ and $2< \alpha <3$ respectively. The NTWS family seems to be the best tool for this differentiation, and the ${S}_\text{W}$ has the best performance in this interval.

\item \textit{fractional Gaussian noise, $-1<\alpha < 1$:}

The TLWS seems inadequate to describe this range---note that $S^\text{(TL)}_\text{W}(12,-1) < S^\text{(TL)}_\text{W}(12,3)$. The $S_\text{W}$ is the best suited to describe these noises, since it is monotonically decreasing and presents a range of variation $\Delta S_\text{W} = 0.698$. While the $\widetilde{S}_\text{W}$ confuses between noises coming from short- or long-memory processes, $-1<\alpha <0$ and $0< \alpha<1$ respectively. It has its maximum at $\alpha = 0$ (white noise).
\end{itemize}

\section{Conclusions}\label{sec:con}

We have introduced exact theoretical expressions for the wavelet entropies associated to fGn, $-1<\alpha<1$. In particular, the range $-1<\alpha<0$, to our knowledge, has never been studied.

We have shown that, at least to characterize fBm's and fGn's processes, the NTWS family seems to be a better quantifier than TLWS. In particular, the $\widetilde{S}_\text{W}$ fulfils all the requirements for a correct description of the overall $\alpha$-range: has its maximum at the white noise, differentiates between noises and processes, and has the maximum range of variation, $\Delta \widetilde{S}_\text{W} = 0.827$. Nevertheless, the $S_\text{W}$ is best suited to discern between different fBm processes. Finally, in the $\alpha > 0$ case, an inverse dependence on $N$ is observed: the NTWS family increases its performance as $N$ increases and the TLWS improves its performance as $N$ decreases. Although the NTWS family always improves with N for any $\alpha$ value.

The procedure outlined in Sec.~\ref{ssec:we} can be followed to build new probability distributions associated to wavelet resolution levels. The weight of each resolution level could be modified according to the requirements of the physical problem under study.


\ack        
 
This work was partially supported by Consejo Nacional de Investigaciones Cient\'{\i}ficas y T\'ecnicas (PIP 5687/05, CONICET, Argentina) and Pontificia Universidad Cat\'olica de Valpara\'iso (Project No. 123.781/2005, PUCV, Chile). DGP and OAR are very grateful to Prof. Dr. Javier Martin\'ez-Mardones 
for his kind hospitality at Instituto de F\'isica, Pontificia Universidad Cat\'olica de Valpara\'iso, Chile, where part of this work was done.

\section*{APPENDIX}

Following the methodology described in  P\'erez \etal~\cite{paper:perez2005} let us take as the signal the noise $s(t) = W^H(t,\omega)$---$\omega$ is fixed and represents one element of the statistic ensemble and it will be omitted hereafter. Now, using the chaos expansion described in Ref.~\cite{paper:perez2004}, any fractional Gaussian noise can be written as
\begin{equation}
W^H(t) = \sum^\infty_{n=1}  M_H\xi_n(t)\,\mathcal{H}_{\epsilon_n}(\omega),
\label{eq:chaos}
\end{equation}
where $\{\xi_n\}_{n\in\N}$ are the Hermite functions, and the operator $M_H$ is defined as follows
\begin{equation}
\widehat{M_H \phi}(\nu)= c_H \abs{\nu}^{1/2-H} \widehat{\phi}(\nu),
\label{eq:H-operator}\end{equation}
where the hat stands for the Fourier transform, $c_H^2= \Gamma(2H+1)\sin (\pi H)$, and $\phi$ is any function  in $L^2(\Re)$.

Given the orthonormal wavelet basis $\{2^{j/2}\psi(2^{j} \cdot - k)\}_{j,k\in\mathbbm{Z}} =\{\psi_{j,k}\}_{j,k\in\mathbbm{Z}}$, we obtain 
\begin{equation}
C^{W^H}_j(k) = \langle W^H, \psi_{j,k}\rangle = \sum^\infty_{n=1}  \langle M_H\xi_n, \psi_{j,k}\rangle\mathcal{H}_{\epsilon_n}(\omega).
\end{equation}
Now we are free to work with the individual coefficients
\begin{equation}
d^H_n(j,k) = \langle M_H\xi_n,\psi_{j,k}\rangle
= c_H \int_\Re \abs{\nu}^{1/2-H} \widehat{\xi}_n(\nu)\, \widehat{\psi}_{j,k}(\nu)\d\nu.
\end{equation}
Since, the Fourier transforms of the Hermite functions and the wavelet are $\widehat{\xi}_n(\nu) = i^{1-n} \xi_n(\nu)$ and $\widehat{\psi}_{j,k}(\nu)= 2^{-j} \exp(-i2^{-j} k\nu )\widehat{\psi}(2^{-j}\nu)$, respectively. The evaluation of the coefficients $d^H_n(j,k)$ is straightforward from their definition:
\begin{equation}
d^H_n(j,k) = c_H i^{1-n}\, 2^{-(H - 1/2)j} \int_\Re \abs{\nu}^{1/2 - H} \Psi(\nu) \xi_n(2^j\nu)\, e^{-i k\nu}\d \nu,
\label{eq:dcoeff}\end{equation}
where $\Psi(\nu)=\widehat{\psi}(\nu)$. 

The chaos expansion in eq.~\eqref{eq:chaos} corresponds to a Gaussian process~\cite{book:holden96}, then under the same procedure used in Ref.~\cite{paper:perez2005} the mean of the squared coefficients results
\begin{align}
\mathbbm{E}\abs{C^{W^H}_j(k)}^2 &=c^2_H  2^{-j(2H -1)  } \int_\Re \abs{\nu}^{-(2H-1)} \abs{\Psi(\nu)}^2\d \nu\nonumber\\
&= 2 \Gamma(2H+1)\sin (\pi H) 2^{-j(2H-1) } \int_0^\infty \nu^{-(2H -1)} \abs{\Psi(\nu)}^2\d \nu,
\label{eq:fgn}
\end{align}
for any $\Psi$ decaying fast enough.

In the case of the fractional Gaussian noises $\alpha = 2H-1$, as opposite to the fractional Brownian motion where $\alpha = 2H + 1$. For the latter we have previously reported~\cite{paper:perez2005} that
\begin{equation}
\mathbbm{E}\abs{C^{B^H}_j(k)}^2 = 2 \Gamma(2H+1)\sin (\pi H) 2^{-j(2H+1)}\int_0^{\infty} \nu^{-(2H +1)} \abs{\Psi(\nu)}^2\d \nu,
\label{eq:fbm}
\end{equation}
for any mother wavelet satisfying $\int_\Re \psi = 0$. Therefore, these two expresions, eqs.~\eqref{eq:fgn} and \eqref{eq:fbm}, can be combined in one written in terms of the power $\alpha$:
\begin{equation}
\mathbbm{E}\abs{C^\alpha_j(k)}^2 = 2\, c^2_H\, 2^{-j\alpha}\int_0^{\infty} \nu^{-\alpha} \abs{\Psi(\nu)}^2\d \nu,
\end{equation}
where $-1<\alpha<1$ or $1<\alpha<3$, and $c_H$ is calculated from the value of $\alpha$.

\bibliography{references}

\begin{thebibliography}{10}
\expandafter\ifx\csname url\endcsname\relax
  \def\url#1{\texttt{#1}}\fi
\expandafter\ifx\csname urlprefix\endcsname\relax\def\urlprefix{URL }\fi

\bibitem{book:mallat1999}
S.~Mallat, A Wavelet tour of signal processing, 2nd Edition, Academic Press,
  1999.

\bibitem{book:daubechies92}
I.~Daubechies, Ten Lectures on Wavelets, SIAM, Philadelphia, 1992.

\bibitem{paper:rosso2001}
O.~A. Rosso, S.~Blanco, J.~Yordanova, V.~Kolev, A.~Figliola, M.~{Sch\"urmann},
  E.~{Ba\c{s}ar}, Wavelet entropy: a new tool for analysis of short duration
  brain electrical signals, J. Neuroscience Method 105 (2001) 65--75.

\bibitem{paper:zunino2004}
L.~Zunino, D.~G. P{\'e}rez, O.~A. Rosso, M.~Garavaglia, Characterization of
  laser propagation through turbulent media by quantifiers based on the wavelet
  transform, Fractals 12~(2) (2004) 223--233.

\bibitem{paper:perez2005}
D.~G. P{\'e}rez, L.~Zunino, M.~Garavaglia, O.~A. Rosso, Wavelet entropy and
  fractional {Brownian} motion time series, accepted to be published in Physica
  A (2005).

\bibitem{paper:blanco98}
S.~Blanco, A.~Figliola, R.~Q. Quiroga, O.~A. Rosso, E.~Serrano, Time-frequency
  analysis of electroencephalogram series. {III. W}avelet packets and
  information cost function, Phys. Rev. E 57 (1998) 932--940.

\bibitem{paper:tavares2005}
D.~M. Tavares, L.~S. Lucena, Entropy analysis of stochastic processes at finite
  resolution, Physica A 357~(1) (2005) 71--78.

\bibitem{paper:shannon48}
C.~E. Shannon, A mathematical theory of communications, Bell Syst. Technol. J.
  27 (1948) 379--423 and 623--656.

\bibitem{paper:mandelbrot1968}
B.~B. Mandelbrot, J.~W.~V. Ness, Fractional {B}rownian motions, fractional
  noises and applications, SIAM Rev. 4 (1968) 422--437.

\bibitem{book:taqqu94}
G.~Samorodnitsky, M.~S. Taqqu, Stable non-Gaussian random processes, Stochastic
  Modeling, Chapman {\&} Hall, London, U.K., 1994.

\bibitem{paper:flandrin1992}
P.~Flandrin, Wavelet analysis and synthesis of fractional {Brownian} motion,
  IEEE Trans. Inform. Theory IT-38~(2) (1992) 910--917.

\bibitem{paper:abry2000}
P.~Abry, P.~Flandrin, M.~S. Taqqu, D.~Veitch, Wavelets for the analysis,
  estimation, and synthesis of scaling data, in: K.~Park, W.~Willinger (Eds.),
  Self-similar Network Traffic and Performance Evaluation, Wiley, 2000.

\bibitem{paper:perez2004}
D.~G. P{\'e}rez, L.~Zunino, M.~Garavaglia, Modeling the turbulent wave-front
  phase as a fractional brownian motion: a new approach, J. Opt. Soc. Am. A
  21~(10) (2004) 1962--1969.
\newline\urlprefix\url{arXiv.org/physics/0403005}

\bibitem{book:holden96}
H.~Holden, B.~{\O}ksendal, J.~Ub{\o}e, T.~Zhang, Stochastic partial
  differential equations: A modeling, white noise functional approach, in:
  Probability and Its Applications, Probability and Its Applications,
  Birkh{\"a}user, 1996.

\end{thebibliography}
\bibliographystyle{elsart-num}

\newpage

\end{document}